\documentclass[12pt, journal]{IEEEtran}
\usepackage[T1]{fontenc}
\usepackage{textcomp}
\usepackage[margin=1in]{geometry}
\usepackage{graphicx}
\usepackage{subcaption}
\usepackage{setspace}
\usepackage{amsmath}
\usepackage{mathtools}

\graphicspath{{figures/}}
\DeclareGraphicsExtensions{.eps, .png}
\begin{document}
\onecolumn
\newcommand{\massbogie}{m_b}
\newcommand{\masscar}{m_c}
\newcommand{\kbogie}{k_b}
\newcommand{\kcar}{k_c}
\newcommand{\cbogie}{c_b}
\newcommand{\ccar}{c_c}
\newcommand{\tf}{\text{H(s)}}
\title{Estimating Passenger Loading on Train Cars Using Accelerometer}
%
%
%

\author{Saurav~R.~Tuladhar,~\IEEEmembership{}
		Peter~Khomchuk,~\IEEEmembership{}        
        and~Siva~Sivananthan\\~\IEEEmembership{ARCON, Waltham, MA}
\thanks{Work performed under Phase I SBIR grant DTRT5717C10231 from the Volpe National Transportation Systems Center, US Department of Transportation}}

%
%

\maketitle

\begin{abstract}
Crowding on train cars is a common problem plaguing the major public transit agencies around the world. On one hand a crowded train car presents a negative experience for the passengers, while on the other hand it indicated inefficiencies in the train system. The Federal Transit Agency is interested in reducing the crowding level on public transit train cars. Automatic passenger counters (APC) are commonly used to count the passengers boarding and alighting the train cars. Advanced APC solutions are available based on EO/IR sensors and visual object detection technology, but are considerably expensive for large scale deployment. This report discusses a low-cost approach to APC by using accelerometer measurements from train car to estimate approximate passenger loading. Accelerometer sensor can measure train car vibration as the train moves along the rail tracks. The train car vibration changes with the passenger loading on the car. Detecting this change in vibration pattern with changing passenger loading level is key to the accelerometer based APC solution. Moreover, accelerometer sensors present a low-cost APC solution compared to existing EO/IR based APCs. This work presents a (i) theoretical model analysis (ii) experimental data driven approach to demonstrate the feasibility of using accelerometer for passenger loading estimation.
\end{abstract}

\pagestyle{plain}
\pagenumbering{arabic}
\setcounter{page}{1}
\section{Introduction}
\label{ch:background}
Crowding in train cars is increasingly a major concern for transit agencies. From the perspective of the passengers and the transit agencies, overcrowding of the train cars has several negative consequences such as: (i) extended duration of passengers boarding and alighting which leads to longer dwell times, (ii) subsequent disruption of the headway and the schedule, and (iii) passenger dissatisfaction (e.g. increased stress and lack of privacy). Transit agencies are interested in developing solutions to mitigate the  negative effects of crowding in their train cars \cite{wmata2014, zheng2013}. 

The first step to minimize the crowding in train car involves measuring the existing crowding level. A common approach to assess crowding level uses automatic passenger counting (APC) devices to measure on board passenger count. In the past, most transit systems relied on manual counting for collecting data and monitoring crowding conditions in the train cars. However, recent advancements in APC technology has indroduced devices that automatically measure the number of passengers entering or exiting a train car \cite{reuter2003passenger}. Such APC devices include infrared or CCTV based counters, treadle mats, and vehicle weighing devices. 

The CCTV video or infra-red (IR) image based APCs are known to provide  the most robust and accurate passenger counts. These vision-based APCs can be up to 97\% accurate and require minimal changes after installation. The CCTV or IR sensors are installed at each door in the train car. They provide directional information, such that passengers entering and leaving a car can be counted separately. The vision-based methods are robust to light and brightness conditions, differences in passengers height and appearance. Since these sensors count only the number of boarding and alighting passengers, and not the total number of passengers in the car at a given time, the train crowding information estimated using these sensors is susceptible to an accumulating error which increases over time, especially after visiting stations with a high passenger flow.

Another set of APC technology is based on directly measuring changes in the weight of a train car due to boarding and alighting passengers. Most of the modern train cars are equipped with electronic weighing sensors which provide information to the train braking system. The data from these sensors can be used to infer passenger load count. Experiments in the Copenhagen rail network show that the weight-based approach provides passenger counts comparable to the infra red-based APCs \cite{Nielsen}. Another approach to weight-based APC uses weight measuring transducers installed on rail lines between two stations. As the train cars pass over the weight sensors, the weights of train cars can be measured. 

The proliferation of WiFi and bluetooth capable devices among the public makes them a suitable candidate for passenger counting. Researchers are developing reliable methods to infer the number of passengers inside a train car using the number of bluetooth devices detected \cite{Maekawa}. However, WiFi and bluetooth based APC technology is not robust enough for systematic deployment for large transit systems \cite{wmata2014}.

In general, the current set of APC technologies are well suited for use in buses to count passengers. Using these APCs on trains may be more difficult due to complications unique to railcar entry. However, some agencies, including Berlin and Hamburg, are using APCs successfully in their trains. Berlin and Hamburg use doorway infrared devices to count passengers in the train cars. 

The data from the APCs are currently being used by transit agencies for various applications such as on-time-performance, revenue analysis, updating schedules and upgrading capacity. But to our knowledge, no transit agency in the US is using the count data to provide crowding information to the waiting passengers.
\section{Accelerometer based APC}
\label{ch:apc}
This section presents the development of the accelerometer based APC solution for measuring passenger count on a train car. The accelerometer based approach method would be an economical solution to APC compared to existing solutions. 

An accelerometer sensor can measure static acceleration like gravity or dynamic acceleration due to the motion or vibration. When properly attached to an object, an accelerometer can be used to measure the dynamic acceleration of the object. The goal of the proposed method is to infer the passenger loading level from the accelerometer measurements on-board a train car.

In classical mechanics, the acceleration of a body subjected to an external input force depends on the body mass. Changing the body mass changes the acceleration given that the same input force is applied. The change in acceleration can be measured using an accelerometer attached to the object. In the same spirit, the accelerometer-based APC is based on the idea that a train carbody vibration should change with changes in the carbody mass due to alighting and boarding passengers. These changes in carbody vibration can be measured using an accelerometer fixed onto the carbody.

The train carbody vibrations impact the ride quality and riders comfort. The allowed levels of vibrations are regulated by ISO 2631-1 standard. The train cars must be designed such that vibrations are minimized in the frequency bands that the passengers are most sensitive to. It was shown that the vertical vibrations have the most impact on riders comfort. A number of theoretical and experimental studies have been performed to measure the impact of vertical train car vibrations on the riders comfort. Among other results the authors in \cite{carlbom2000carbody, nagai2006coupled, tomioka2015experimental} report that the carbody vibration spectrum changes when the number of passengers change, and the effect is most noticeable in the frequency range of the carbody's first flexible vibration mode. Although the results of these studies focus on the ride quality, they confirm our intuition that it is possible to detect changes in the train carbody mass by measuring vertical vibrations.
 
One of the major goals of the Phase I effort is to demonstrate a PoC of using train car accelerometer data to implement an APC. This PoC also marks the first milestone of the Phase I effort. ARCON pursued a two-pronged approach to demonstrate the PoC for the accelerometer based APC:
\begin{itemize}
\item \textbf{Model analysis} This approach used simplified theoretical models for train carbody to analyze the relation between carbody mass and vertical vibration.  
\item \textbf{Data driven analysis} This approach focused on gathering accelerometer data from an operational train car and conducting analysis to establish the relation between passenger loading and the vibration measurements.
\end{itemize}
As a part of the two-pronged approach, ARCON successfully completed a series of theoretical and experimental analyses to establish a correlation between the carbody vertical vibration and the passenger loading. This correlation between the measurement and passenger loading demonstrates the proof-of-concept for using accelerometer data to estimate passenger loading on a train car. The sections below describe two approaches in detail.

\section{Model analysis approach}
A common approach when analyzing a complex system such as a train car is to start with a simplified theoretical model to build an initial insight. Train cars are usually modeled using simple mass-spring-damper systems with varying degrees-of-freedom (DOF). This simplified model allows us to analytically solve for the quantity of interest. For our purpose we are interested in deriving the relation between the train car vertical vibration and its mass. 

\begin{figure}[!ht]
\centering
\includegraphics[width=0.7\textwidth]{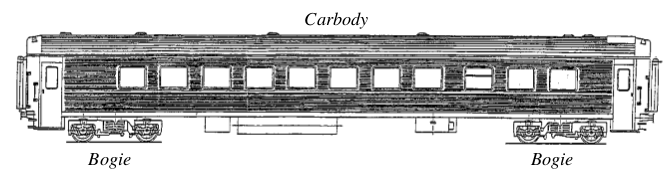}
\caption{Train car consists of a carbody mounted on a pair of bogies.}
\label{fig:carbody}
\end{figure}

Figure~\ref{fig:carbody} shows a typical train car which consists of the carbody mounted on a pair of bogies \cite{carlbom2000carbody,garg2012dynamics}. The passengers board the carbody which is connected to the bogies via a secondary suspension system. Each bogie has a pair of wheels on either side resting on a pair of track rails (not shown in figure). The wheels are connected to the bogie via a primary suspension system. The wheels are assumed to follow the track geometry as the train car moves along the track. Any variation on the track geometry will vertically displace the wheels. This vertical displacement of the wheels are transmitted via the suspension system to the carbody. The vertical vibration of the carbody is a combined effect of the input from the track geometry variation transformed via the suspension systems and the mechanical response of the carbody. The track geometry variations are the primary dynamic inputs to the train car. For a train car moving with a constant velocity, the wheel displacement due to the track geometry variation can be viewed as a time-series input. The input from track geometry variation is commonly modeled as a periodically modulated stochastic process \cite[Section~3.8]{garg2012dynamics}.

The vertical vibration observed in the carbody is a combination of rigid body motion (bounce and pitch) and flexible bending motion. The rigid body motion is what is usually perceived by the train passengers as vibration. It is a slow shaking of the car up and down due to unevenness of tracks. The flexible body vibrations occur since the train car is not an ideal rigid body and it can experience a certain amount of bending and torsion. The vibration motion is spatially distributed along the carbody. The spatial distribution of the vibration can be resolved into different spatial mode shapes. Each mode shape has a different temporal frequency. The rigid modes and flexible modes exists in distinct frequency bands \cite{carlbom2000carbody,tomioka2015experimental}. This spectral separation of vertical vibration modes forms the basis for the spectral analysis of vertical vibration.

Two models were considered for the model based analysis. The first model was a simple 2-degrees-of-freedom (2-DOF) mass-spring-damper system. This model captures only the rigid mode motion, but provides a simple analytical solution for vertical vibration. The second model was a multi-DOF system which captures both rigid modes and flexible modes. The multi-DOF system is a more realistic model of a train car.

\subsection{2-DOF mass-spring-damper model}
\begin{figure}[!ht]
\centering
\includegraphics[width=0.25\textwidth]{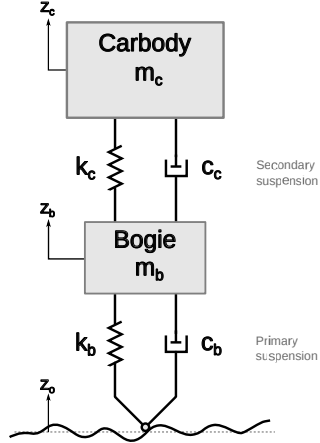}
\caption{2-DOF spring-mass-damper model for train carbody excited by track irregularities.}
\label{fig:twodof_model}
\end{figure}

Figure~\ref{fig:twodof_model} shows a 2-DOF mass-spring-damper system used as a model for the train car. The two mass units correspond to the carbody mass, $m_c$, and the bogie mass, $m_b$. The secondary suspension is modeled by a spring-damper system with spring constant, $k_c$, and damper coefficient, $c_c$. The bogie is assumed to have a single point contact with the track via the primary suspension. The primary suspension is modeled by the spring-damper system with spring constant, $k_b$, and damper coefficient, $c_b$. The entire model is assumed to be moving at a constant velocity, $v$. As the model moves forward, the primary suspension point contact is displaced vertically following the track geometry. The displacement of the point contact is denoted by $z_o$. This displacement due to the track irregularities is the only input applied to the model. The resulting displacement of the carbody mass and the bogie mass are denoted by $z_c$ and $z_b$ respectively.

\begin{figure}[!ht]
\centering    
\includegraphics[width=0.7\textwidth]{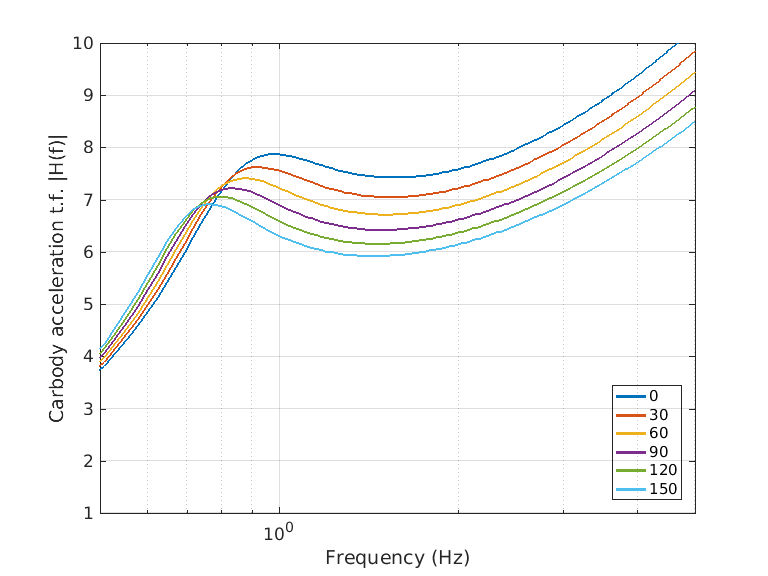}        
\caption{Carbody vertical displacement acceleration output to track input transfer function.}
\label{fig:twodof_tf}
\end{figure}

The 2-DOF model has two observable displacements and a single input in the form of the base displacement. Hence, the equations of motions are,
 \begin{align}
 \label{eq:2dof-eqmo}
 \masscar \ddot{z_c} + \ccar(\dot{z_c} - \dot{z_b}) + \kcar(z_c - z_b) =& 0 \nonumber \\ 
 \massbogie \ddot{z_b} + \ccar(\dot{z_b} - \dot{z_c}) + \kcar(z_b - z_c) + \cbogie\dot{z_b} + \kbogie z_b =& \cbogie\dot{z_o} + \kbogie z_o
 \end{align}
Our interest is in understanding how the carbody vertical displacement responds to the base excitation from the track. Using Eq.~\ref{eq:2dof-eqmo}, the transfer function between the carbody displacement, $z_c$, and the base displacement, $z_o$, is
 \begin{align}
 \label{eq:2dof_tf}
 \tf = \frac{\ccar\cbogie s^2 + (\ccar\kbogie + \cbogie\kcar)s + \kcar\kbogie}{\massbogie\masscar s^4 + (\ccar\massbogie + (\ccar-\cbogie)\masscar)s^3 + (\ccar\cbogie + \kcar\massbogie + (\kbogie - \kcar)\masscar)s^2 + (\ccar\kbogie + \cbogie\kcar)s + \kcar\kbogie}
 \end{align}
Eq.~\ref{eq:2dof_tf} relates the carbody displacement to the base excitation input and clearly shows that the carbody displacement response depends on the carbody mass. Figure~\ref{fig:twodof_tf} shows the transfer functions computed using the parameters in Table~\ref{tbl:2dof_parameters}. Each color corresponds to a different passenger loading level indicated in the legend. Each passenger is assumed to weigh 70kgs. Increasing the passenger count effectively increases the carbody mass and the transfer function also changes. Considering the transfer function at 1~Hz, it is observed that the transfer function magnitude decreases with increase in the passenger loading.

\begin{table}[!ht]
\centering
\caption{2-DOF model simulation parameters.}
\label{tbl:2dof_parameters}	
	\begin{tabular}{|r|l|}
    	\hline
         carbody mass($m_b$) & 38000kg \\
         bogie mass ($m_t$) & 3000kg\\
         primary suspension stiffness ($k_c$) & 0.78 $\times 10^6$~N/m\\
         secondary suspension stiffness ($k_b$) & 0.55 $\times 10^6$~N/m\\
         primary suspension damping ($c_c$) & 30 $\times 10^6$~Ns/m\\
         secondary suspension damping ($c_b$) & 60 $\times 10^3$~Ns/m\\
         mass per person & 70kg \\
         \hline
    \end{tabular}	
\end{table}

\begin{figure}[!ht]
\centering
\includegraphics[width=0.7\textwidth]{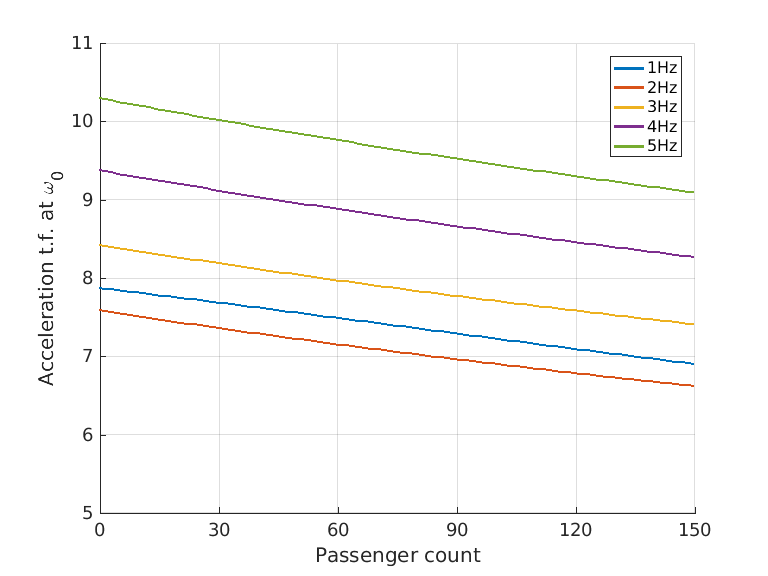}
\caption{Transfer function at a fixed frequency vs passenger loading}
\label{fig:twodof_fxfreqtf}   
\end{figure}

Figure~\ref{fig:twodof_fxfreqtf} shows how the transfer function magnitude at a particular frequency changes with increase in passenger loading. The x-axis corresponds to the passenger count and the y-axis corresponds to the transfer function magnitude ($|H(f_0)|$) evaluated at a fixed frequency $f_0$. Each curve corresponds to a different value of $f_0$ indicated in the legend. In all five frequency values considered in Figure~\ref{fig:twodof_fxfreqtf}, the transfer function magnitude decreases with increase in the passenger count. Clearly the change in carbody mass impacts the vertical vibration response of the carbody. This change in the carbody vibration due to change in passenger loading is the basis of the proposed accelerometer based APC. 

\subsection{Multi-DOF model}
\label{sec:mdof}
\begin{figure}[!ht]
\centering
\includegraphics[width=0.9\textwidth]{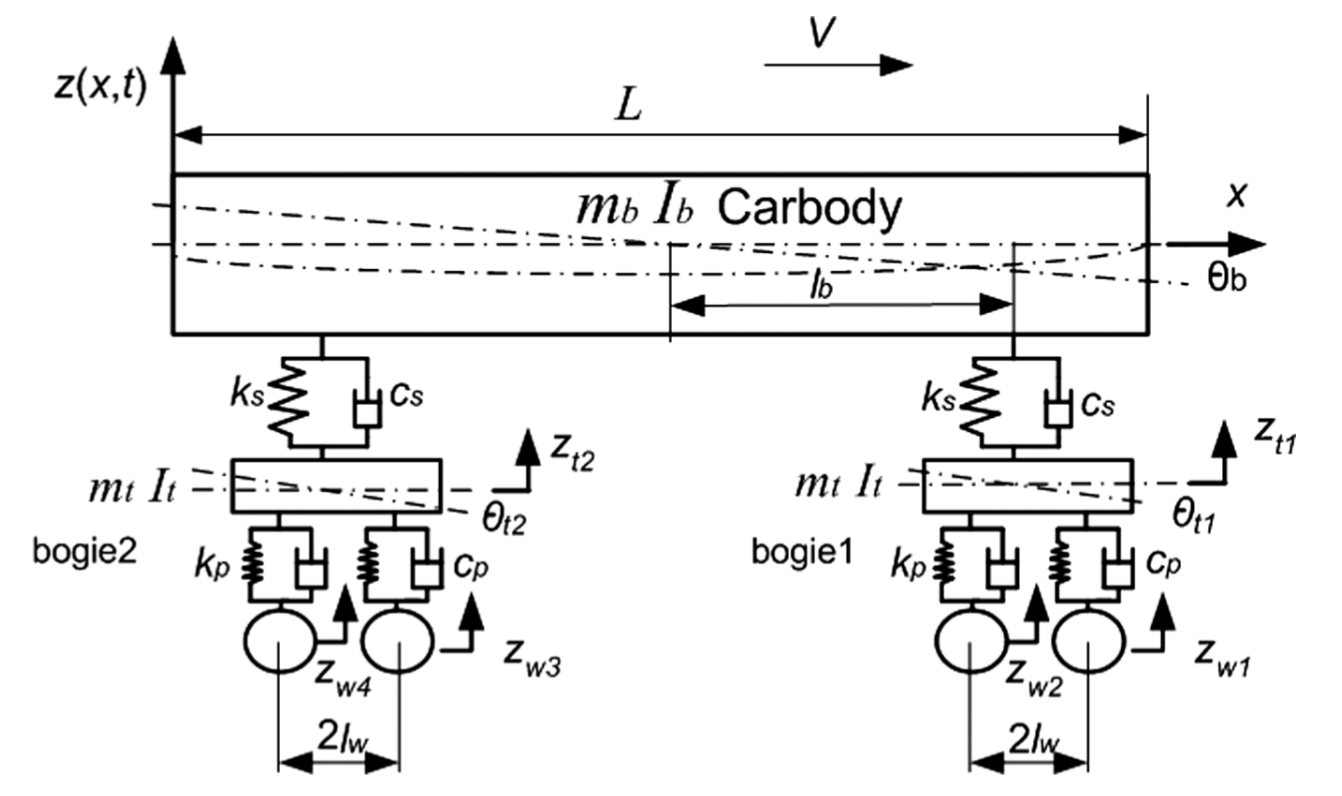}
\caption{Multi-DOF model of train carbody mounted on two bogie units \cite{zhou2009influences}.}
\label{fig:mdof_model}
\end{figure}

Multi-DOF models are used for higher fidelity representation of the carbody response \cite{garg2012dynamics}. Multi-DOF models can capture vibration due to both rigid body motion and flexible motion. For the purpose of Phase I task, we considered the multi-DOF model shown in Figure~\ref{fig:mdof_model}, in which only vertical vibrations are considered. In this model, the carbody is assumed to be a simple uniform Euler-Bernoulli beam supported by the secondary suspensions on the bogies. It is assumed that when the train car moves on the track, the wheels follow the track irregularities \cite{zhou2009influences}.

\begin{table}[!ht]
\centering
\caption{Mutli-DOF model simulation parameters.}
\label{tb:sim-params}	
	\begin{tabular}{|r|l|}
    	\hline
         carbody mass($m_b$) & 28000\,kg \\
         carbody pitch inertia & 1.3 $\times 10^6$\,kg m\textsuperscript{2} \\
         carbody bending stiffness & 4.987 $\times 10^9$\,kN/m\textsuperscript{th}\\
         carbody internal damping coefficient& 1.936 $\times 10^6$\,kN s/m\textsuperscript{th}\\
         carbody length ($L$) & 24.5\,m\\
         bogie mass ($m_t$) & 2500\,kg\\
         bogie pitch inertia & 1500\,kg$\cdot$m\textsuperscript{th} \\
         half distance between bogies ($l_b$) & 8.75\,m\\
         half wheelbase & 1.25\,m\\
         primary suspension stiffness ($k_p$) & 2.4 $\times 10^6$\,N/m\\
         secondary suspension stiffness ($k_s$) & 0.5 $\times 10^6$\,N/m\\
         primary suspension damping ($c_p$) & 30 $\times 10^6$\,Ns/m\\
         secondary suspension damping ($c_s$) & 60 $\times 10^3$\,Ns/m\\
         
         mass per passenger & 70\,kg \\
         train speed & 30 mph (50\,km/h) \\
         track excitation intensity & 1.080 $\time 10^{-6}$\\
         cut-off frequency $\Omega_c$ & 0.8246\,cycles/m\\
         cut-off frequency $\Omega_r$ & 0.0206\,cycles/m \\
         \hline
    \end{tabular}	
\end{table}

The details of a multi-DOF model according to \cite{zhou2009influences} and the derivation of a  corresponding equation of motion is provided in Appendix \ref{appendix:multi_dof}. In matrix form the equation is
\begin{align}
\label{eq:mdof}
\mathbf{M}\ddot{y} + \mathbf{C}\dot{y} + \mathbf{K}y = \mathbf{D_w}z_w + \mathbf{D_{dw}}\dot{z}_w,
\end{align}
where $y$ is a vector of observed parameters, one of which is the vertical displacement $z(x, t)$. $\mathbf{M}$, $\mathbf{C}$ and $\mathbf{K}$ are inertia, damping and spring stiffness matrices respectively, and $\mathbf{D_w}$ and $\mathbf{D_{dw}}$ are the track displacement and velocity matrices. The vertical displacement $z(x, t)$ is a function of space (along the carbody length) and time. 

\begin{figure}[!ht]
\centering
\includegraphics[width=0.65\textwidth]{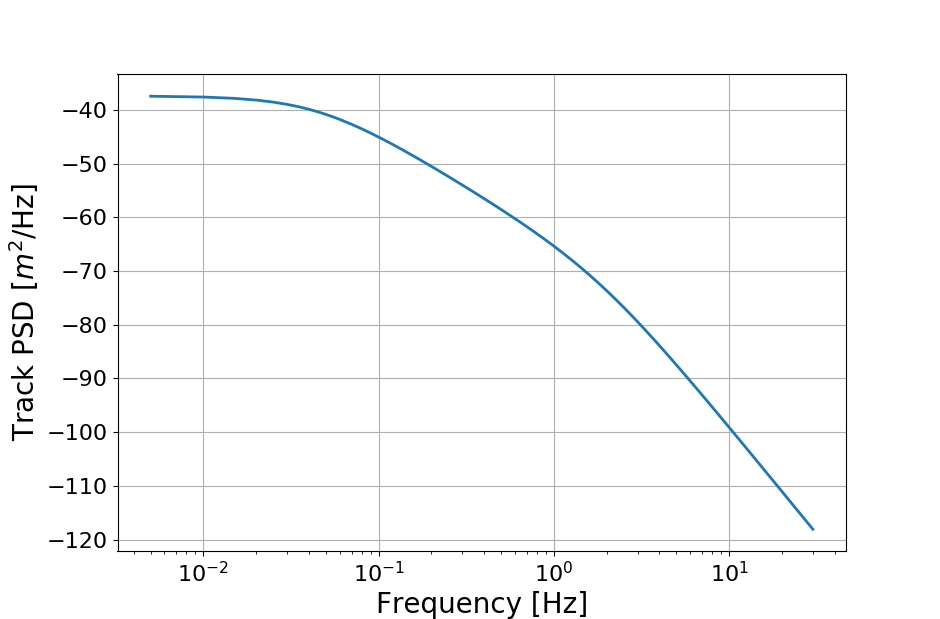}
\caption{Power spectral density of track geometry variation.}
\label{fig:track_psd}
\end{figure}

Vertical track irregularities, $z_w(t)$, is the input to the system described by Eq.~(\ref{eq:mdof}). The output of the system is $z(x, t)$, the vertical displacement of the car as a function of position inside the car. A common approach to modeling the vertical track irregularities is to consider them as a random process with a known power spectral density (PSD). The track PSD depends on the train speed and track quality. For our simulations we use the model described in \cite{hamid1983analytical}:
\begin{equation}
S_{track}(\omega) = \frac{A_v, \Omega_c^2 V^3}{\omega^4 + (\Omega^2 + \Omega^2) V^2 \omega^2 + \Omega_r^2\Omega_c^2 V^4}
\end{equation}
where $\omega = 2 \pi f$ is the circular frequency measured in radians per second, $V$ is the train speed in m/s, $A_v$, $\Omega_c$, and $\Omega_r$ are the track excitation intensity, and the two cut-off frequencies respectively that control the track quality. Parameters $A_v$, $\Omega_c$, and $\Omega_r$ are defined in literature for different track qualities, the values used in our simulation are given in Table \ref{tb:sim-params} and correspond to track grade 2 according to American Railway standard. The corresponding track PSD is shown in Figure \ref{fig:track_psd}.

\begin{figure}[!ht]
    \centering
    \begin{subfigure}[b]{0.9\textwidth}
        \includegraphics[width=\textwidth]{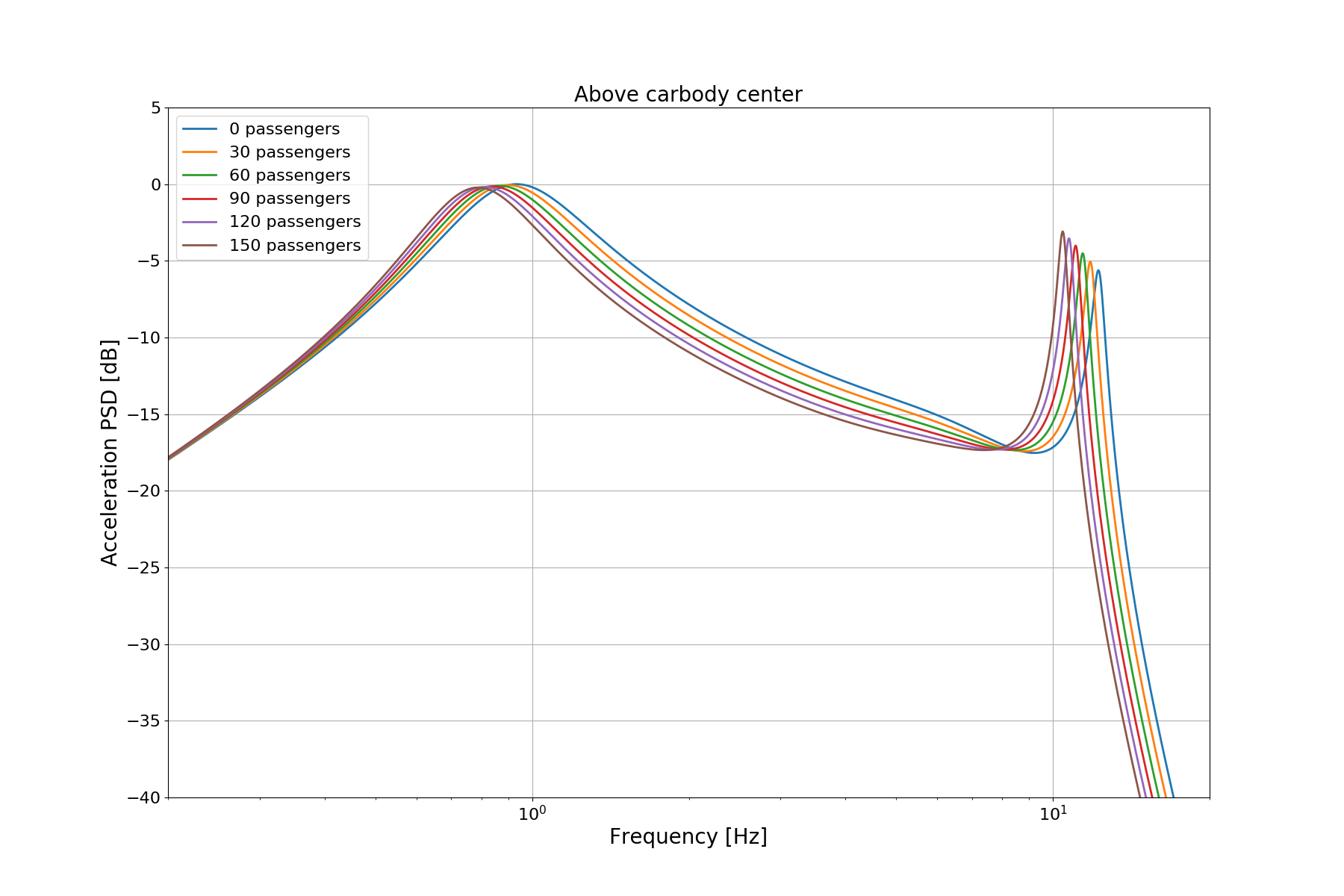}
        \caption{Above carbody center}
        \label{fig:mdof_psd_center}
    \end{subfigure}
    \hfill
      
    \begin{subfigure}[b]{0.9\textwidth}
        \includegraphics[width=\textwidth]{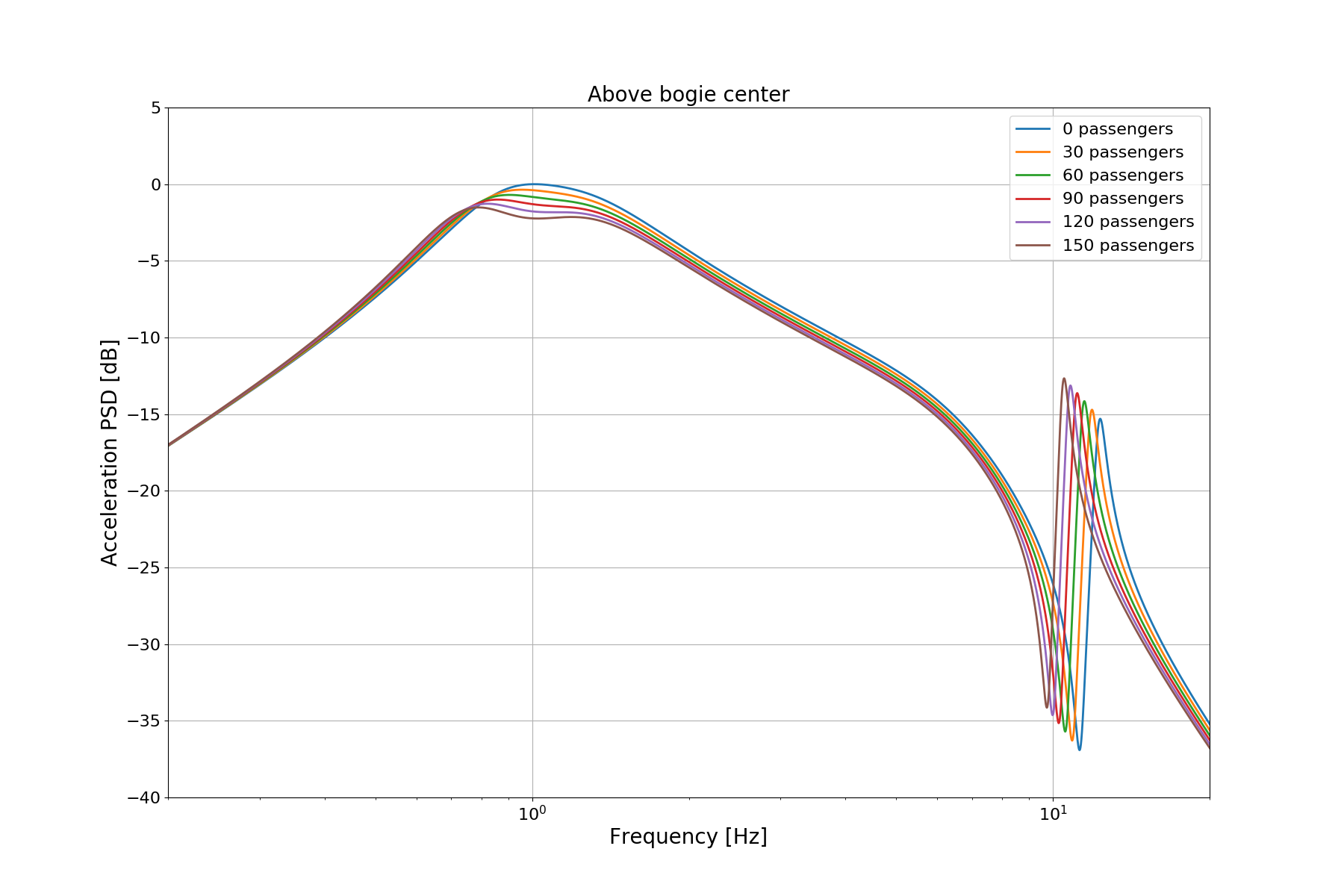}
        \caption{Above bogie}
        \label{fig:mdof_psd_bogie}
    \end{subfigure}
    \caption{Vertical displacement acceleration PSD for the multi-DOF model.}    
    \label{fig:mdof_psd}
\end{figure}

Our analysis looks at the spectral properties of the carbody vertical displacement acceleration at different points along the carbody. Figure~\ref{fig:mdof_psd_center} shows the vertical vibration acceleration PSD  at the center of the carbody. The PSD exhibits prominent peaks around the 1~Hz and 10~Hz regions after which there is a sharp decay. The peak around 1~Hz corresponds to the rigid mode vibration and the peak around 10~Hz corresponds to the flexible mode vibration. Again, each curve in Figure~\ref{fig:mdof_psd_center} represents the acceleration PSD for different passenger loading level indicated in the legend. Comparing the curves shows that the spectral peaks shift towards lower frequency as the passenger loading increases. Also, for a fixed frequency values between 1~Hz and 5~Hz, the magnitude of the PSD decreases for higher passenger loading. Similarly, Figure~\ref{fig:mdof_psd_bogie} shows the vertical vibration acceleration PSD observed above the bogie. The PSD shape is similar to  Figure~\ref{fig:mdof_psd_center}. Again the acceleration PSD changes with increase in the passenger loading.

Clearly there is a correlation between the passenger loading and the spectral properties of the vertical acceleration of the carbody. This change in acceleration PSD should be observable using an accelerometer suitably mounted on board of a carbody. We should be able to infer the current passenger loading from the accelerometer measurements, based on the above results from the multi-DOF train car model.

\section{Data driven approach}
The accelerometer based APC will use measurements from the accelerometer sensors suitably placed inside the train car-body. As a part of the Phase I tasks, we planned to collect accelerometer data on operational train cars and perform analysis to see how the data relates to the results from model analysis. In order to conduct the data collection, ARCON implemented an accelerometer datalogger using low-cost open source solutions.

\subsection{Raspberry Pi based datalogger}
\begin{figure}[!ht]
\centering
\includegraphics[width=0.8\textwidth]{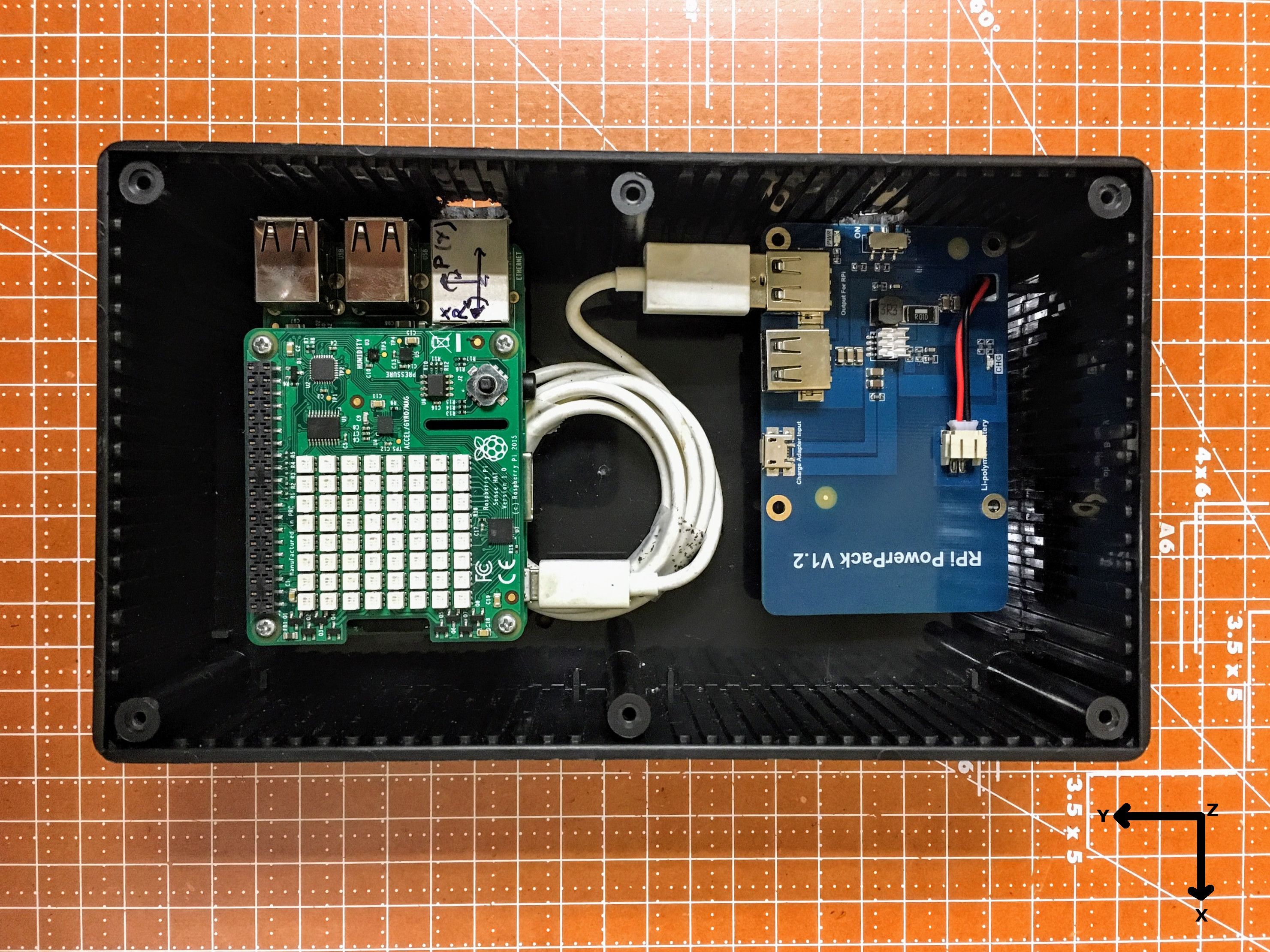}
\caption{Top view of the internal arrangement of the Raspberry Pi based datalogger prototype.}
\label{fig:logger-topview}
\end{figure}

The datalogger prototype is implemented using the Raspberry Pi 3 single-board computer with an attached Sense HAT module \cite{raspi3, sensehat}. The Sense HAT module is equipped with a 9 degrees-of-freedom (DOF) IMU sensor. The IMU sensor is capable of recording 3-channel acceleration, angular rate and magnetic field. The Raspberry Pi with the Sense HAT module is powered using a rechargeable battery unit. The Rapsberry Pi with the Sense HAT module and the battery module is placed inside an enclosure creating a self-contained portable datalogger unit. Figure~\ref{fig:logger-topview} shows the internal arrangement of the datalogger unit. We have prepared two such datalogger units for the initial data collection on MBTA subway trains. 

The datalogger software operates on the Raspberry Pi and records the IMU measurements at a preset sample rate. At each timestamp, the datalogger records 3D accelerations, 3D angular rates, 3D magnetic field values and the fused 3-D orientation. The recorded values are saved to an on-board SD card and retrieved later for post analysis.

\subsection{Data collection}
ARCON collected data on the subway trains operated by the Massachusetts Bay Transportation Authority (MBTA) \cite{mbta}. We completed several data collection sessions on the MBTA Red Line subway line. Two team members carrying the datalogger units rode the Red Line train cars across several stations \cite{mbta}. The team members placed the datalogger units on the train car floor and initiated data logging. The datalogger units were placed approximately at the center of the train car. Upon departing each station, our team members recorded the approximate passenger count. At the end of each session, the collected data was analyzed.

During each new data collection session, adjustments were made to the data collection procedure based on the lessons learned from the previous session. However after four sessions further data collection was halted for the following reasons,
\begin{itemize}
\item Lack of control over the data collection environment: The team members were riding the subway as normal passengers and hence lacked any control over the data collection environment. It was not possible to guarantee that the logger was placed at the same location on each session. 
\item Improper mounting of the sensors: Accelerometer sensor needs to be properly mounted onto the object whose measurement is desired. But during our data collection sessions, the datalogger was simply placed on the floor without any arrangement to hold it fixed at one position. Without proper mounting, we are unable to ascertain the fidelity of the acceleration measurements recorded. 
\end{itemize}

\subsection{Accelerometer Data Example}
The MBTA Red Line subway line runs in the northwest-to-southeast direction across the Greater Boston region. The northern terminus is the Alewife station and the trains moving away from the Alewife station are considered to be in the \emph{Inbound} direction where as the trains moving towards the Alewife station are considered to be in the \emph{Outbound} direction. 

On August 8, 2017 we conducted four data collection sessions in the inbound and outbound direction between the Alewife station, Davis station, Porter station and the Harvard station. This approach produced three data records per session for inbound and outbound direction respectively. Each data record corresponds to train traveling between two adjacent stations. Table~\ref{tbl:pax_count} lists the approximate passenger count observed during the ride through four stations. Table~\ref{tbl:pax_count_outbound} lists the approximate passenger count observed during the ride in the outbound direction. The first row lists the adjacent stations traversed during the train ride and the first column indicates the approximate starting time for each session.

\begin{table}[!ht]
\centering
\caption{Passenger count during inbound trips}
\label{tbl:pax_count}
\begin{tabular}{|l|l|l|l|l}
\cline{1-4}
         			& Alewife-Davis & Davis-Porter & Porter-Harvard &  \\ \cline{1-4}
Session1 (9:00AM) 	& 20       & 60       & 80       &  \\ \cline{1-4}
Session2 (9:30AM)	& 6        & 12       & 20       &  \\ \cline{1-4}
Session3 (10:30AM)	& 19       & 22       & 25       &  \\ \cline{1-4}
Session4 (11:00AM)	& 10       & 20       & 30       &  \\ \cline{1-4}
\end{tabular}
\end{table}

\begin{table}[!ht]
\centering
\caption{Passenger count during outbound trips}
\label{tbl:pax_count_outbound}
\begin{tabular}{|l|l|l|l|l}
\cline{1-4}
         			& Harvard-Porter & Porter-Davis & Davis-Alewife &  \\ \cline{1-4}
Session1 (9:15AM) 	& 5       & 3       & 5       &  \\ \cline{1-4}
Session2 (9:45AM)	& 45        & 25      & 8       &  \\ \cline{1-4}
Session3 (10:50AM)	& 18       & 13       & 8       &  \\ \cline{1-4}
Session4 (11:30AM)	& 20       & 20       & 4       &  \\ \cline{1-4}
\end{tabular}
\end{table}

\begin{figure}[!ht]
\centering
\includegraphics[width=\textwidth]{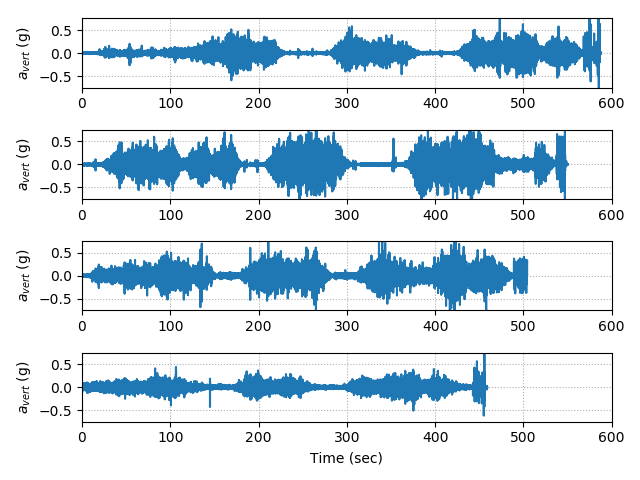}
\caption{Vertical vibration acceleration measured on MBTA Redline subway train car traveling in the inbound direction from the Alewife station to the Harvard station. Each panel corresponds to a separate data collection session along same set of station}
\label{fig:mbta_accelz}
\end{figure}

\begin{figure}[!ht]
\centering
\includegraphics[width=\textwidth]{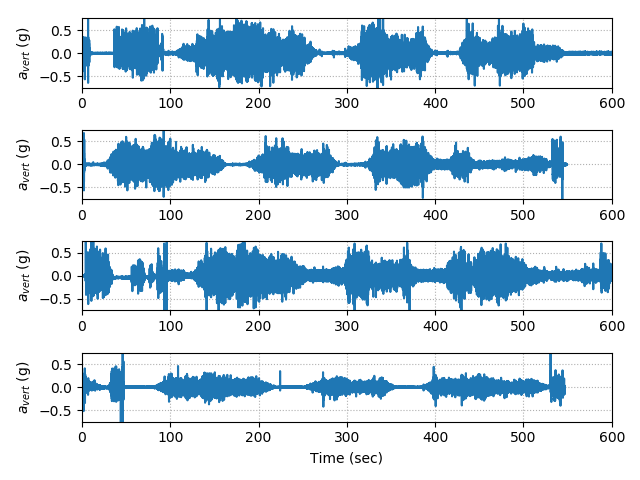}
\caption{Vertical vibration acceleration measured on MBTA Redline subway train car traveling in the outbound direction from the Harvard station to the Alewife station. }
\label{fig:mbta_accelz_outbound}
\end{figure}

Figure~\ref{fig:mbta_accelz} shows the time-series plots of the raw vertical acceleration measurements collected during each session in the inbound direction. The four panels in Figure~\ref{fig:mbta_accelz} correspond to the four sessions mentioned in Table~\ref{tbl:pax_count}. Each time-series plot has segments when the measured acceleration is approximately zero and segments when the measured acceleration is visibly much higher. The latter set of segments correspond to the interval when the train is in motion. As mentioned earlier, when the train is in motion, the track irregularities drive the vertical vibration observed on the train car. In the same spirit, Figure~\ref{fig:mbta_accelz_outbound} shows the time-series plots of the raw vertical acceleration measurements collected during each session in the outbound direction.

\begin{figure}[!ht]
    \centering
    \begin{subfigure}[b]{0.8\textwidth}
        \includegraphics[width=\textwidth]{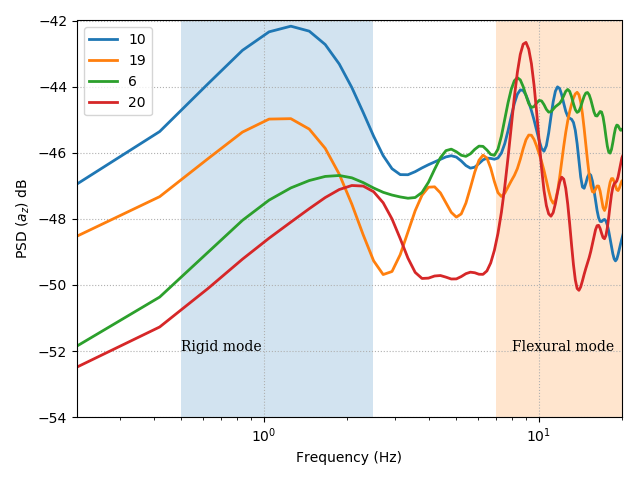}
        \caption{Alewife to Davis station }
        \label{fig:psd_station1}
    \end{subfigure}
    \hfill
    \begin{subfigure}[b]{0.785\textwidth}
        \includegraphics[width=\textwidth]{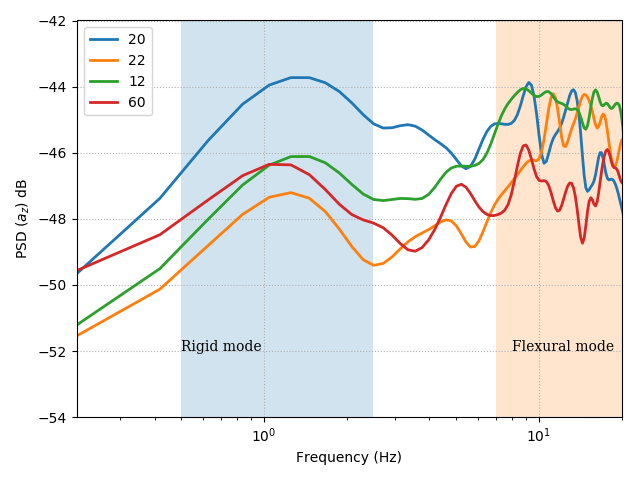}
        \caption{Davis to Porter station}
        \label{fig:psd_station2}
    \end{subfigure}    
    \newpage
    \caption{PSD of vertical acceleration measured on an MBTA Redline subway train. Each curve corresponds to the passenger loading level indicated in the legend.}
	\label{fig:mbta_accelz_psd}
\end{figure}

Figure~\ref{fig:mbta_accelz_psd} shows the PSD of the vertical acceleration measured when traveling from the Alewife to the Porter station (top panel) and traveling from the Porter station to the Davis station. In each panel, the four PSD curves correspond to the data records from the four sessions in Table~\ref{tbl:pax_count}. The legend indicates the passenger count associated with each PSD curve. Comparing the PSD curves in  Figure~\ref{fig:mbta_accelz_psd} and the PSD curves of Figure~\ref{fig:mdof_psd_center} shows the similarity between the acceleration PSD derived from the multi-DOF model and the PSD computed from real data. The PSD in Figure~\ref{fig:mbta_accelz_psd} also has spectral peaks at frequencies corresponding to the rigid mode motion and the flexible mode motion. However, the correlation between the passenger count and the PSD curve is not observable as was the case with the multi-DOF model. It should be noted that the current analysis is based on a very limited set of real data collected from the MBTA subway. Lack of sufficient amount of data impedes us from a statistically valid assessment of the relation between passenger loading and the spectral properties of the vertical acceleration observations of train carbody.

\subsection{Issues}
Several issues were encountered during the Phase I effort to establish the proof-of-concept for accelerometer based APC. The major issues are discussed below.

In the model based analysis approach, the multi-DOF train car model requires values for the train car parameters and track geometry spectral parameters listed in Table~\ref{tb:sim-params}. For the analysis presented in Section \ref{sec:mdof}, the parameter values were referenced from \cite{zhou2009influences}. However for a higher fidelity model of a real train car, we need matching parameter values to be used with the model. Such parameter values may be obtained only from manufacturers or operators.
A higher fidelity model will eventually allow us to derive a more accurate algorithm to estimate passenger loading from the accelerometer measurements.

In the data driven approach, our efforts were severely limited by the lack of access to the MBTA train infrastructure. The accelerometer data logger should be securely attached to the train carbody to obtain useful acceleration measurements. When securely attached, we can be certain that (i) all  measurements are from same location on the train car and (ii) improper mounting has minimal impact on the measured data quality. Without any access to the MBTA infrastructure, we are able to obtain only limited data measurements. Moreover since the data logger was simply placed on the carbody floor, we cannot be certain that the accelerometer measurements are not affected by spurious vibrations. 

\appendix
\section{Multi-DOF model of train carbody vibrations}
\label{appendix:multi_dof}
In \cite{zhou2009influences} authors model a train carbody as a simple uniform Euler-Bernouli beam. This is a half-car model, i.e. only a half of the train car is considered with two bogies and a pair of wheels at each bogie. The carbody of mass, $m_b$, and length, $L$, is supported by a set of secondary suspensions with spring stiffness, $k_s$, and damping coefficient, $c_s$, per bogie as shown in Figure \ref{fig:mdof_model}. The secondary suspensions are attached to the bogies which are modeled as solid (non-flexible) bodies with equal masses, $m_t$. Bogies can only experience bounce and pitch displacements. A pair of wheels is attached to each bogie through primary suspension with spring stiffness and damping coefficient, $k_p$, and, $c_p$, respectively. It is assumed that the wheels cannot jump. Thus, wheels can only move up and down in correspondence to the irregularities in the rail track. From the beam theory the equation for Euler-Bernouli beam deflections is
\begin{equation}
\label{eq:euler_bernouili}
EI \dfrac{ \partial^4 z(x,t)}{\partial x^4} + \mu I \frac{\partial^5 z(x,t)}{\partial t \partial x^4} + \rho \frac{\partial ^ 2 z(x,t)}{\partial t^2} = P_1 \delta(x-l_1) + P_2\delta(x-l_2),
\end{equation}
where 
$E$ is an elastic modulus of the carbody, $I$ is an equivalent beam moment of inertia, $\mu$ is a structural damping coefficient, $\rho = m_b/L$ is a mass per unit length, $z(x,t)$ is a carbody displacement at the location $x$ and time $t$,
$P_1$ and $P_2$ are the forces acted by the secondary suspension on the carbody at positions $l_1$ and $l_2$ respectively (Figure \ref{fig:mdof_model})
\begin{eqnarray}
\label{eq:force1}
P_1 = -k_s[z(l_1, t) - z_{t1}] - c_s[\dot{z}(l_1,t)-\dot{z}_{t1}],\\
\label{eq:force2}
P_2 = -k_s[z(l_2, t) - z_{t2}] - c_s[\dot{z}(l_2,t)-\dot{z}_{t2}],
\end{eqnarray}
and $z_{t1}$ and $z_{t2}$ are the vertical displacements of the bogies.

The partial differential equation in (\ref{eq:euler_bernouili}) can be solved by using the variables separation method. The carbody deflections, $z(x,t)$, can be represented using functions, $Y_i(x)$, that describe spatial shapes of the deflection modes and thus depend only on the spatial coordinate $x$, and functions, $q_i(t)$, known as modal coordinates, that describe temporal variations and depend only on time variable $t$
\begin{equation}
\label{eq:displacement}
z(x,t) = z_b(t) + \left(\frac{L}{2} - x \right)\theta_b(t) + \sum_{i=3}^n Y_i(x)q_i(t),
\end{equation}
here $z_b(t)$ denotes the bounce rigid mode of the carbody (i.e. $q_1(t) = z_b(t)$ and $Y_1(x) = 1$ ), $\theta_b(t)$ is the carbody pitch angle ($q_2(t) = \theta_b(t)$ and $Y_2(x) = L/2 - x$), and $n$ is total number of the considered modes. It can be shown that the higher order ($i>2$) modes are the flexible modes that have the following spatial shapes
\begin{equation}
Y_i(x) = \cosh\beta_ix+\cos\beta_ix - \frac{\cosh\lambda_i - \cos\lambda_i}{\sinh\lambda_i - \sin\lambda_i}\left(\sinh\beta_ix + \sin\beta_ix\right),
\end{equation}
where $\lambda_i$ and $\beta_i$ must satisfy
\begin{eqnarray}
1-\cosh\lambda_i\cos\lambda_i = 0 \textrm{, } \beta_i = \frac{\lambda_i}{L}.
\end{eqnarray}
Solutions for modal coordinates, $q_i(t)$, can be obtained by substituting (\ref{eq:displacement}) into (\ref{eq:euler_bernouili}) and integrating along the lengths of the carbody in order to eliminate the spatial variable $x$. Since the modal shape functions, $Y_i(x)$, are orthogonal to each other for different $i$, we obtain
\begin{align}
\label{eq:carbody_modes1}
m_b&\ddot{z}_b(t) =P_1+P_2,\\
I_b&\ddot{\theta}_b(t) =P_1\left(\frac{L}{2}-l_1\right)+P_2\left(\frac{L}{2}-l_2\right),\\
\ddot{q}_i&(t)  +\frac{\mu I \beta_i^4}{\rho}\dot{q_i(t)} + \frac{EI\beta_i^4}{\rho}q_i(t) = \frac{Y_i(l_1)}{m_b}P_1 + \frac{Y_i(l_2)}{m_b}P_2, i=3,4,\ldots,n \label{eq:higher_order_modes}
\end{align}
were $I_b$ is the carbody pitch inertia.

Equation (\ref{eq:higher_order_modes}) is a second order differential equation that describes a driven harmonic oscillator (a second order system). Hence this equation for the $i^{\rm th}$ mode can be rewritten in terms of the natural frequency, $\omega_i$, and the damping ratio, $\xi_i$, of the corresponding harmonic oscillator
\begin{equation}
\ddot{q}_i(t) +2\xi_i\omega_i\dot{q_i(t)} + \omega_i^2q_i(t) = \frac{Y_i(l_1)}{m_b}P_1 + \frac{Y_i(l_2)}{m_b}P_2 \textrm{, } i=3,4,\ldots,n 
\end{equation}
where 
\begin{equation}
\omega_i^2 = \frac{EI\beta_i^4}{\rho}\textrm{, } \xi_i = \frac{\mu I \beta_i^4}{2\rho\omega_i}.
\end{equation}
Further, the forces in (\ref{eq:force1}) and (\ref{eq:force2}) can be rewritten using (\ref{eq:displacement}) as
\begin{align}
\label{eq:force1_2}
P_1 = & -k_s\left[z_b(t)+\left(\frac{L}{2}-l_1\right)\theta_b(t) + \sum_{i=3}^n Y_i(l_1)q_i(t)-z_{t1}\right]\nonumber\\
&-c_s\left[\dot{z}_b(t) + \left(\frac{L}{2}-l_1\right)\dot{\theta}_b(t) + \sum_{i=3}^n Y_i(l_1)\dot{q}_i(t) - \dot{z}_{t1}\right],\\
\label{eq:force2_2}
P_2 = & -k_s\left[z_b(t)+\left(\frac{L}{2}-l_2\right)\theta_b(t) + \sum_{i=3}^n Y_i(l_2)q_i(t)-z_{t2}\right]\nonumber\\
&-c_s\left[\dot{z}_b(t) + \left(\frac{L}{2}-l_2\right)\dot{\theta}_b(t) + \sum_{i=3}^n Y_i(l_2)\dot{q}_i(t) - \dot{z}_{t2}\right].
\end{align}

The equations for bogie bounce and pitch follow from Newton's second law
\begin{align}
\label{eq:bogie1}
m_t\ddot{z}_{t1} =& -k_s(z_{t1}-z(l_1,t))-c_s(\dot{z}_{t1}-\dot{z}(l_1,t))-k_p(z_{t1}-l_w\theta_{t1}-z_{w1}) \nonumber\\
&-c_p(\dot{z}_{t1}-l_w\dot{\theta}_{t1}-\dot{z}_{w1}) -k_p(z_{t1}+l_w\theta_{t1}-z_{w2})-c_p(\dot{z}_{t1}+l_w\dot{\theta}_{t1}-\dot{z}_{w2}),\\
I_t\ddot{\theta}_{t1} =&l_wk_p(z_{t1}-l_w\theta_{t1}-z_{w1}) + l_w c_p(\dot{z}_{t1}-l_w\dot{\theta}_{t1}-\dot{z}_{w1} ) \nonumber \\
&-l_w k_p(z_{t1}+l_w\theta_{t1}-z_{w2})-l_w c_p(\dot{z}_{t1}+l_w\dot{\theta}_{t1}-\dot{z}_{w2}),\\
m_t\ddot{z}_{t2} =& -k_s(z_{t2}-z(l_2,t))-c_s(\dot{z}_{t2}-\dot{z}(l_2,t))-k_p(z_{t2}-l_w\theta_{t2}-z_{w3}) \nonumber\\
&-c_p(\dot{z}_{t2}-l_w\dot{\theta}_{t2}-\dot{z}_{w3}) -k_p(z_{t2}+l_w\theta_{t2}-z_{w4})-c_p(\dot{z}_{t2}+l_w\dot{\theta}_{t2}-\dot{z}_{w4}),\\
I_t\ddot{\theta}_{t2} =&l_wk_p(z_{t2}-l_w\theta_{t2}-z_{w3}) + l_w c_p(\dot{z}_{t2}-l_w\dot{\theta}_{t2}-\dot{z}_{w3} ) \nonumber \\
&-l_w k_p(z_{t2}+l_w\theta_{t2}-z_{w4})-l_w c_p(\dot{z}_{t2}+l_w\dot{\theta}_{t2}-\dot{z}_{w4}),
\label{eq:bogie4}
\end{align}
where $I_t$ is the boige pitch inertia, $\theta_{t1}$ and $\theta_{t2}$ are the pitch angles of the first and the second bogie respectively, $l_w$ is a half of the bogie wheel base (Figure \ref{fig:mdof_model}), and $z_{w1},\ldots,z_{w4}$ are the vertical track irregularities experienced by the four wheels. These irregularities are considered to be the inputs to the system.

Further by substituting equations (\ref{eq:force1_2}) and (\ref{eq:force2_2}) into (\ref{eq:carbody_modes1})-(\ref{eq:higher_order_modes}) and then rewriting these equations together with (\ref{eq:bogie1})-(\ref{eq:bogie4}), the following matrix equation can be obtained
\begin{align}
\label{eq:mat_form}
\mathbf{M}\ddot{y} + \mathbf{C}\dot{y} + \mathbf{K}y = \mathbf{D_w}z_w + \mathbf{D_{dw}}\dot{z}_w,
\end{align}
where $y$ is a vector that contains all unknown variables 
\begin{equation*}
y = [z_b(t), \theta_b(t), q_3(t),\ldots,q_n(t),z_{t1}(t),z_{t2}(t),\theta_{t1}(t),\theta_{t2}(2)]^T,
\end{equation*}
and $z_w$ is a vector of input variables
\begin{equation*}
z_w = [z_{w1},z_{w2},z_{w3},z_{w4}]^T
\end{equation*}
In (\ref{eq:mat_form}) $\mathbf{M}$, $\mathbf{C}$, $\mathbf{K}$ are the inertia, damping, and stiffness matrices respectively, $\mathbf{D_w}$, $\mathbf{D_{dw}}$ are the track displacement and the velocity input matrices. These matrices contain the coefficients that appear next to the corresponding variables with same order of derivatives in equations (\ref{eq:carbody_modes1})-(\ref{eq:higher_order_modes}) and (\ref{eq:bogie1})-(\ref{eq:bogie4}).

In order to analyze the frequency characteristics of the carbody vibrations, the transfer function can be obtain by taking the Laplace transform of both sides of the equation in (\ref{eq:mat_form}) 
\begin{align}
\mathbf{M}&\mathcal{L}\left\{\ddot{y}\right\} + \mathbf{C}\mathcal{L}\left\{\dot{y}\right\} + \mathbf{K}\mathcal{L}\left\{y\right\} = \mathbf{D_w}\mathcal{L}\left\{z_w \right\} + \mathbf{D_{dw}}\mathcal{L}\{\dot{z}_w\mathcal\} \nonumber\\
\mathbf{M}&s^2\mathcal{Y}(s) + \mathbf{C}s\mathcal{Y}(s) + \mathbf{K}\mathcal{Y}(s) = \mathbf{D_w}s\mathcal{Z}_w(s) + \mathbf{D_{dw}}\mathcal{Z}_w(s)\nonumber\\
\mathcal{Y}(s)& \left[\mathbf{M}s^2+ \mathbf{C}s + \mathbf{K}\right] = \mathcal{Z}_w(s)\left[\mathbf{D_w}s + \mathbf{D_{dw}}\right]\nonumber\\
\mathcal{H}(s)& = \frac{\mathcal{Z}_w(s)}{\mathcal{Y}(s)}=\left[\mathbf{M}s^2+ \mathbf{C}s + \mathbf{K}\right]^{-1}\left[\mathbf{D_w}s + \mathbf{D_{dw}}\right]
\end{align}
where $\mathcal{H}(s)$ is a system transfer function, and $s$ is a complex number frequency parameter.
\subsection*{Acknowledgement}
ARCON Corporation would like to thank the Volpe National Transportation Systems Center for the SBIR grant which gave us the opportunity to work on the proposed ideas for the Crowding Information Collection and Dissemination (CICD) system. We are grateful to Dr.~Nazy Sobhi for being the technical point-of-contact for the Phase I effort. Dr.~Sobhi's guidance, feedbacks and encouragements during monthly meetings helped to steer this project in the right direction.
Finally, we would like to acknowledge Dr.~Laurel Paget-Seekins,  the Director of Strategic Initiatives at the Massachusetts Bay Transportation Authority (MBTA). We thank Dr.~Seekins for sharing her insights into the practical aspects of the train crowding problem. 

\bibliographystyle{ieeetr}
\bibliography{cicd} 
\end{document}